\documentclass[aps,prl,twocolumn,floatfix,superscriptaddress,showpacs,amsfonts
,amssymb,amsmath,preprintnumbers]{revtex4-1}
\usepackage{bm}
\usepackage{epsfig}
\usepackage{graphicx}
\usepackage{color}
\usepackage{subfigure}

\newcommand{\eq}[1]{Eq.~(\ref{#1})}
\newcommand{\eqs}[2]{Eqs.~(\ref{#1}--\ref{#2})}

\newcommand{\be}{\begin{equation}}
\newcommand{\ee}{\end{equation}}
\newcommand{\beq}{\begin{equation}}
\newcommand{\eeq}{\end{equation}}

\newcommand\bea{\begin{eqnarray}}
\newcommand\eea{\end{eqnarray}}

\def\llambda{\lambda/L}

\begin{document}
\title{Role of magnetic reconnection in MHD turbulence}
\author{N.\ F.\ Loureiro}
\affiliation{Plasma Science and Fusion Center, Massachusetts Institute of Technology, Cambridge MA 02139, USA}
\author{S. Boldyrev}
\affiliation{Department of Physics, University of Wisconsin at Madison, Madison, WI 53706, USA}
\affiliation{Space Science Institute, Boulder, Colorado 80301, USA}

\date{\today}

\begin{abstract}
The current understanding of MHD turbulence envisions turbulent eddies which are anisotropic in all three directions. In the plane perpendicular to the local mean magnetic field, this implies that such eddies become current-sheet-like structures at small scales. We analyze the role of magnetic reconnection in these structures and conclude that reconnection  becomes important at a scale $\lambda\sim L S_L^{-4/7}$, where $S_L$ is the outer-scale ($L$) Lundquist number and $\lambda$ is the smallest of the field-perpendicular eddy dimensions. 
This scale is larger than the scale set by the resistive diffusion of eddies, therefore implying a fundamentally different route to energy dissipation than that predicted by the Kolmogorov-like phenomenology. 
In particular, our analysis predicts the existence of the sub-inertial, reconnection interval of MHD turbulence, with the Fourier energy spectrum $E(k_\perp)\propto k_\perp^{-5/2}$, where $k_\perp$ is the wave number perpendicular to the local mean magnetic field. The same calculation is also performed for high (perpendicular) magnetic Prandtl number plasmas ($Pm$), where the reconnection scale is found to be $\lambda/L\sim S_L^{-4/7}Pm^{-2/7}$.
\end{abstract}

% insert suggested PACS numbers in braces on next line

\maketitle

%***********************************************************************************

\paragraph{Introduction.}
Turbulence is a defining feature of magnetized plasmas in space and astrophysical environments, which are almost invariably characterized by very large Reynolds numbers. 
The solar wind~\cite{bale_measurement_2005}, the interstellar medium~\cite{armstrong_density_1981,armstrong_electron_1995}, and accretion disks \cite[e.g.,][]{balbus1998,walker_etal2016} are prominent examples of plasmas dominated by turbulence, where its detailed understanding is almost certainly key to addressing long-standing puzzles such as electron-ion energy partition, cosmic ray acceleration, magnetic dynamo action, and momentum transport. 

Weak collisionality implies that kinetic plasma physics is required to fully describe turbulence in many such environments~\cite{schekochihin_astrophysical_2009}. 
However, turbulent motions at scales ranging from the system size to the ion kinetic scales, 
an interval which spans many orders of magnitude, should be accurately described by magnetohydrodynamics (MHD). 

The current theoretical understanding of MHD turbulence largely rests on the ideas that were put forth by Kolmogorov and others to describe turbulence in neutral fluids (the K41 theory of turbulence~\cite{kolmogorov_local_1941}), and then adapted to magnetized plasmas by Iroshnikov and Kraichnan~\cite{iroshinkov_turbulence_1963,kraichnan_inertial_1965} and, later, Goldreich and Sridhar (GS95) ~\cite{goldreich_toward_1995}. 
Very briefly, one considers energy injection at some large scale, $L$, (the forcing, or outer, scale) which then cascades to smaller scales through the inertial range where, by definition, dissipation is negligible and throughout which, therefore, energy is conserved. At the bottom of the cascade is the dissipation range, where the gradients in the flow are sufficiently large for the dissipation to be efficient.

Turbulence in magnetized plasmas fundamentally differs from that in neutral fluids due to the intrinsic anisotropy introduced by the magnetic field. GS95 suggests this leads to turbulent eddies which are longer in the direction aligned with the local field than in the direction perpendicular to it. 
The relationship between the field-parallel and perpendicular dimensions is set by critical balance: $V_{A,0}/\ell \sim v_\lambda /\lambda$, where $V_{A,0}$ is the Alfv\'en velocity based on the background magnetic field $B_0$, $\ell$ and $\lambda$ are, respectively, the field-aligned and field-perpendicular dimensions of the eddy, and $v_\lambda$ is the velocity perturbation at that scale.

More recently, it was argued~\cite{boldyrev_spectrum_2006} that the GS95 picture of turbulence needs to be amended to allow for angular alignment of the velocity and magnetic field perturbations at scale $\lambda$. 
As a result, eddies are also anisotropic in the plane perpendicular to the local magnetic field, being thus characterized by three scales: $\ell$, along the field, and $\lambda$ and $\xi$, perpendicular to the field.
Although the precise structure of MHD turbulence remains an open research question, observational and numerical evidence in support of 3D anisotropic eddies has since been reported~\cite{podesta_scale_2009,chen_3D_2012, mallet_measures_2016}.

A particularly interesting feature of 3D anisotropic eddies is that they can be thought of as current sheets of 
thickness $\lambda$ and length $\xi$ in the field-perpendicular plane (with $\xi \gg \lambda$). Following the standard Kolmogorov-like arguments, one would then conclude that the inertial interval ends when the scale $\lambda$ becomes comparable to the dissipation scale.  Below this scale the energy is strongly dissipated in the current sheets. Currently available numerical simulations indicate that a considerable fraction of small-scale current sheets look like sites of magnetic reconnection \cite[e.g.,][]{zhdankin_etal2013,zhdankin_etal2014}.  We note that such a dissipation channel is not a feature of the GS95 model, which predicts filament-like eddies at small scales. 

In this Letter we propose that at sufficiently large magnetic Reynolds numbers, the route to energy dissipation in MHD turbulence is fundamentally different from that envisioned in the Kolmogorov-like theory. This hapens since the anisotropic, current-sheet-like eddies become the sites of magnetic reconnection {\em before} the formal Kolmogorov dissipation scale is reached. This Letter presents the first analytical attempt to quantify this phenomenon and to  characterize the role of reconnection in MHD turbulence.\\

\paragraph{Background.}
The 3D anisotropic eddies that we envision are depicted in Fig. 2 of Ref.~\cite{boldyrev_spectrum_2006}. 
We will characterize them by the smallest of their field-perpendicular dimensions, $\lambda$; other quantities of interest to us here are related to $\lambda$ as follows~\cite{boldyrev_spectrum_2006}:
\bea
\label{eq:xi}
\xi\sim L (\llambda)^{3/4},\\
\label{eq:ell}
\ell \sim L (\llambda)^{1/2},\\
\label{eq:b}
b_\lambda\sim B_0 (\llambda)^{1/4},\\
\label{eq:v}
v_\lambda \sim v_0 (\llambda)^{1/4},\\
\label{eq:tau}
\tau \sim \ell/V_{A,0}\sim\lambda^{1/2} L^{1/2} / V_{A,0},\\
\label{eq:va}
V_{A,\lambda}\sim V_{A,0}(\llambda)^{1/4},
\eea
where $b_\lambda$ and $v_\lambda$ are the magnetic  field and velocity perturbations at scale $\lambda$, $\tau$ the eddy turn-over-time, and the other quantities have already been introduced~\footnote{
It is important to clarify that although we will adopt the scalings~(\ref{eq:xi}--\ref{eq:tau}), the critical assumption underlying our calculation is only that the eddies have different field-perpendicular dimensions such that they become current sheets. If future theoretical considerations lead to different scalings, our calculation here can be easily modified to accommodate them.}.
We will, for simplicity, consider the case where the turbulence is critically balanced at the outer scale such that the outer-scale Lundquist number, $S_L \equiv L V_{A,0}/\eta$ is comparable to the outer-scale Reynolds number, $R_m \equiv L V_0/\eta$. We also introduce the Lundquist number associated with scale $\lambda$, $S_\lambda \equiv \lambda V_{A,\lambda}/\eta$.~\footnote{We assume that magnetic and velocity fluctuations at the outer scale $L$ are not aligned, and are on the same order, $v_0\sim V_{A,0}$.}

A lower bound on the dissipation scale can be obtained from these scalings by equating $\tau$ with the eddy resistive diffusion time, $\lambda^2/\eta$. This yields
\be
\label{eq:lambda_dissipation}
\llambda\sim S_L^{-2/3}\sim R_m^{-2/3}.
\ee
{~}\\

\paragraph{Magnetic Reconnection.}
Let us begin by observing that the aspect ratio of an eddy in the perpendicular direction is 
\be
\label{eq:aspect_ratio}
\xi/\lambda \sim (L/\lambda)^{1/4},
\ee
i.e., it increases as $\lambda\rightarrow 0$. So, in the perpendicular plane, eddies become ever more elongated current sheets as $\lambda$ gets smaller. This is qualitatively different from the GS picture, where both field-perpendicular dimensions are the same, and so the eddy tends to a point in the perpendicular plane as $\lambda\rightarrow 0$.

It is therefore natural to ask at what scale (i.e., aspect ratio) does reconnection of these current sheets (eddies) become an important effect, if ever. If it does, it should leave a well defined signature in both the magnetic and kinetic energy spectra. It may not, however,  correspond to the energy dissipation scale, since reconnection, in addition to dissipating magnetic energy, also accelerates flows. \\

\paragraph{Sweet-Parker reconnection of eddies.}
The simplest estimate that can be done for eddy reconnection  stems from the Sweet-Parker model~\cite{parker_sweet_1957,sweet_neutral_1958}, according to which the scale $\lambda$ at which an eddy would reconnect is given by
\be
\label{eq:SP}
\lambda/\xi \sim S_\xi^{-1/2},
\ee
where $S_\xi = \xi V_{A,\lambda} /\eta$ is the Lundquist number pertaining to a current sheet of length $\xi$, at scale $\lambda$, defined with the Alfv\'en velocity based on the perturbed magnetic field at that scale, \eq{eq:b}.
Using \eqs{eq:xi}{eq:va} above, one finds that Eqs. (\ref{eq:SP}) and (\ref{eq:lambda_dissipation}) are equivalent statements. 

This important observation immediately points to the problem with the Kolmogorov-like transition to the dissipation regime. A robust conclusion of the past decade of reconnection research is that Sweet-Parker current sheets above a certain critical aspect ratio, corresponding to a Lundquist number $ S_c\sim 10^4$, are violently unstable to the formation of multiple magnetic islands, or plasmoids (see~\cite{loureiro_magnetic_2016} for a recent review). One straightforward implication of this instability~\cite{loureiro_instability_2007,samtaney_formation_2009,bhattacharjee_fast_2009,loureiro_plasmoid_2013} is that the Sweet-Parker current sheets cannot be formed in the first place~\cite{loureiro_instability_2007,uzdensky_magnetic_2014,pucci_reconnection_2014,comisso_general_2016}. We now demonstrate that the MHD turbulent cascade will be affected by this instability {\em before} it has a chance to form Sweet-Parker current sheets at small scales, thus qualitatively changing the route to energy dissipation in MHD turbulence. \\

\paragraph{Dynamic reconnection onset and eddy disruption by the tearing instability.}

Consider, as an example, a current sheet of length $L$ and width $a$ that is forming in time (its aspect ratio $L/a$ is increasing at a certain rate)~\cite{uzdensky_magnetic_2014}. 
We know that if it were to reach the Sweet-Parker aspect ratio, $L/a\sim S_L^{1/2}$ it would be unstable to the plasmoid instability, with an instability growth rate $\gamma L/V_{A,0}\sim S_L^{1/4}\gg1$. 
This implies that there would necessarily be an earlier time, when the aspect ratio of the forming current sheet was not yet quite so large, when it would become marginally stable to the tearing (plasmoid) instability. 
As the aspect ratio continues to increase, the tearing instability becomes stronger, overcoming the current sheet formation rate. The linear and, importantly, nonlinear evolution of the tearing instability in this forming sheet enables the computation of the moment of time, and all the current sheet properties at that time, when the magnetic island(s) resulting from the tearing instability become as large the current sheet itself, disrupting its further formation~\cite{uzdensky_magnetic_2014}.

We argue that these ideas remain adequate in the context of statistically steady state MHD turbulence that we are concerned with here, except that the role of time in the above discussion is now played by the scale $\lambda$. In other words, we ask at what scale $\lambda$ is the aspect ratio of the eddies, $\xi/\lambda$, such that their tearing instability is strong enough to warrant significant reconnection in one eddy turn over time.

The tearing instability has two well-known regimes, FKR (small tearing mode instability parameter, $\Delta'$)~\cite{FKR} and Coppi (large $\Delta'$)~\cite{coppi_resistive_1976}. 
The $N=1$ mode, related to the tearing perturbation wavenumber through $k/2\pi = N/\xi$, is the most unstable mode until it transitions into the Coppi regime. This happens at the scale that satisfies
\be
(\xi/\lambda)\, S_\lambda^{-1/4} \sim 1,
\ee
yielding the {\it transition scale} for the $N=1$ mode
\be
\lambda_{tr,1}/L \sim S_L^{-4/9}.
\ee
In other words, if $\lambda>\lambda_{tr,1}$, the most unstable mode in the current sheet is an FKR mode; if the opposite is true, it is instead a Coppi mode which is the most unstable.

The {\it critical scale} for any mode $N$, $\lambda_{cr,N}$, is the scale at which the growth rate of that mode matches the eddy turn over time at that scale, given by \eq{eq:tau}. 
For the $N=1$ mode while in the FKR regime, the growth rate is $\gamma_{1}^{\rm FKR} \sim \xi^{2/5}V_{A,\lambda}^{2/5}\lambda^{-2}\eta^{3/5}$.
The equation $\gamma_{1}^{\rm FKR}\tau\sim 1$ therefore yields
\be
\lambda_{cr,1}/L \sim S_L^{-6/11}.
\ee
We see that $\lambda_{cr,1} < \lambda_{tr,1}$, implying that the modes that will become critical are not FKR modes, but rather Coppi modes. For these modes the largest growth rate is $\gamma_{\rm max}^{\rm Coppi}\sim \tau_{A,\lambda}^{-1} S_\lambda^{-1/2}$, where $\tau_{A,\lambda}\equiv \lambda/V_{A,\lambda}$, corresponding to a mode number $N_{\rm max}^{\rm Coppi} \sim \xi / \lambda\, S_\lambda^{-1/4}$. 
The criticality condition $\gamma_{\rm max}^{\rm Coppi}\tau\sim 1$ now yields
\be
\label{eq:lambda_coppi}
\lambda_{cr}^{\rm Coppi}/L \sim S_L^{-4/7},
\ee
corresponding to a mode number (number of magnetic islands, or plasmoids)
\be
N_{\rm max}^{\rm Coppi}\sim S_L^{1/14}.
\ee

Coppi modes undergo X-point collapse (a loss of equilibrium that happens on the Alfv\'enic timescale at scale $\lambda$, $\tau_{A,\lambda}$)~\cite{waelbroeck_onset_1993,loureiro_X-point_2005} immediately as they become nonlinear.
Therefore,  \eq{eq:lambda_coppi} identifies the scale at which reconnection becomes dynamically relevant to the turbulence: the islands born of current sheets at that scale will very quickly grow to become as wide as $\lambda_{cr}^{\rm Coppi}$, disrupting the current sheet (i.e., the eddy) in which they formed~\cite{uzdensky_magnetic_2014}.

A final observation is that the width of the inner boundary layer of the tearing instability corresponding to this most unstable mode is
\be
\delta_{in,\rm max}^{\rm Coppi}/L\sim S_L^{-9/14}.
\ee
Whether this scale is larger or smaller than kinetic scales in the plasma at hand (the ion gyroscale, ion-acoustic scale, or the ion skin depth) decides the adequateness, or lack thereof, of the MHD description of turbulence at these scales. The extension of the calculation presented here to the kinetic regime would follow the same conceptual guidelines: given eddy scalings at scales below the ion Larmor radius (or skin depth), and the tearing mode scalings in such collisionless regimes, one needs to compute the scale at which the tearing growth rate becomes comparable to the eddy turn over time. This assumes that the eddies remain anisotropic in the field-perpendicular direction at the kinetic scales, such that they can be thought of as current sheets in that plane. The theory of kinetic-scale turbulence, however, has not been developed in sufficient detail yet to conduct this analysis.\\

\paragraph{Large magnetic Prandtl number.}
The calculation above can be straightforwardly repeated for cases in which the magnetic Prandtl number, $Pm\equiv \nu_\perp/\eta$, is large. We are referring to the perpendicular viscosity, not the parallel one; on this matter, the reader is referred to the discussion in section II B of \cite{loureiro_plasmoid_2013}. The perpendicular magnetic Prandtl number that we consider here is $Pm \sim (m_i/m_e)^{1/2}\beta_i$, and it can be large in astrophysical plasmas. 

The scalings for the linear tearing mode in the small and large $\Delta'$ regimes at high Prandtl number were derived in~\cite{porcelli_viscous_1987} and are conveniently summarized in~\cite{loureiro_plasmoid_2013}. 
Since this calculation is entirely similar to the one in the previous section, we limit ourselves to stating the main results.
For the $N=1$ mode in the $Pm\gg 1$ regime, the transition scale is 
$\lambda_{tr,1}/L \sim S_L^{-4/9} Pm^{2/9}$, 
whereas the critical scale is 
$\lambda_{cr,1}/L \sim S_L^{-8/15}Pm^{-2/15}$. 
Clearly $\lambda_{cr,1}\ll \lambda_{tr,1}$ implying, as above, that the modes that will become critical are Coppi modes. 
The critical scale is now
\be
\label{eq:lambda_coppi_Pm}
\lambda_{cr}^{\rm Coppi}/L \sim S_L^{-4/7} Pm^{-2/7},
\ee
corresponding to mode number $N_{\rm max}^{\rm Coppi}=S_L^{1/14}Pm^{15/56}$. The inner boundary layer now scales as
$\delta_{in,\rm{max}}^{\rm Coppi}/L \sim S_L^{-9/14}Pm^{5/56}$.

We see that \eq{eq:lambda_coppi_Pm} yields a smaller scale than its inviscid counterpart, \eq{eq:lambda_coppi}. 
This makes intuitive sense: viscosity slows down the Coppi modes; 
as such, the tearing and turbulence timescales can only match at a scale $\lambda$ smaller (and, therefore, larger current sheet aspect ratio, $\xi/\lambda$) than in the absence of viscosity.

The nonlinear evolution of large $Pm$ tearing modes is less well understood than the low $Pm$ case. 
However, since X-point collapse is an ideal loss of equilibrium, it should not be significantly affected by large viscosity. Thus, as in the inviscid case, \eq{eq:lambda_coppi_Pm} identifies the scale at which the eddies become disrupted by magnetic islands.\\

\paragraph{Spectrum of turbulence below the reconnection scale.}
We now address the spectrum below the reconnection scale identified by ~\eq{eq:lambda_coppi}, or \eq{eq:lambda_coppi_Pm} for $Pm \gg 1$ plasmas.  As the island chain becomes nonlinear and undergoes $X$-point collapse, new current sheets will form between each two plasmoids. These may themselves be unstable to plasmoid formation, and so on.  Assuming one can apply here what is known from the dynamics of large Lundquist number reconnecting systems (see, e.g.~\cite{uzdensky_fast_2010, fermo_statistical_2010,loureiro_magnetic_2012, huang_distribution_2012,huang_plasmoid_2013,loureiro_magnetic_2016}), the final state is one where there is a distribution of plasmoid sizes, whose dynamics is dictated by advection out of the current sheet, coalescence, and generation of new plasmoids. This can be viewed as a new sub-inertial-range interval of turbulence, which may be characterized by its own power spectrum~\cite{barta_spontaneous_2011,huang_turbulent_2016,beresnyak_three_2013}. 

In order to derive the spectrum in this interval, which we call the ``reconnection interval", we first note that we expect such plasmoids of many different sizes to be separated from each other by Sweet-Parker current sheets of a length, $L_c$, such that their aspect ratio is marginally stable to plasmoid formation, $\sim S_c^{1/2}$~\cite{uzdensky_fast_2010}, where $S_c=L_c V_{A,\lambda_{cr}^{\rm Coppi}}/\eta$ is the critical Lundquist number, $S_c \sim 10^4$. 
The rationale is that current sheets longer than $L_c$, and therefore larger values of the Lundquist number, are unstable to plasmoid formation; if, on the other hand, they are shorter than $L_c$, they will be stretched to that length by differential background flows~\cite{uzdensky_fast_2010}. 

The thickness of these critical current sheets is estimated as:
\be
\label{delta_c}
\delta_c \sim L_c S_c^{-1/2} \sim \lambda_{cr}^{\rm Coppi} S_{\lambda_{cr}^{\rm Coppi}}^{-1} S_c^{1/2}, 
\ee
where $S_{\lambda_{cr}^{\rm Coppi}} = \lambda_{cr}^{\rm Coppi} V_{A,\lambda_{cr}^{\rm Coppi}}/\eta$.
Using \eq{eq:lambda_coppi}, we thus obtain
\be
\label{eq:diss_scale}
\delta_c/L \sim S_c^{1/2} S_L^{-6/7}.
\ee

These critical current sheets are the structures where ohmic and viscous dissipation is happening~\cite{loureiro_magnetic_2012}. It is reasonable to assume that Eq.~(\ref{eq:diss_scale}) sets the dissipation scale, that is, the scale below which the reconnection interval is ultimately terminated by the dissipation. Assuming that the energy spectrum in this interval follows a power law, we write it in the form:
\be
E(k_\perp)\propto k_0^{-3/2}\left(k_\perp/k_0\right)^{-\alpha},
\label{eq:spectrum}
\ee
where $k_0\sim 1/\lambda_{cr}^{\rm Coppi}$ is the wavenumber corresponding to the reconnection scale~(\ref{eq:lambda_coppi}), where the reconnection-interval spectrum matches the inertial-interval spectrum of MHD turbulence $E(k_\perp)\propto k_\perp^{-3/2}$. The power-law spectrum (\ref{eq:spectrum}) extends up to the wavenumber $k_*\sim 1/\delta_c$ corresponding to the dissipation scale~(\ref{eq:diss_scale}), after which it is expected to decline fast.  

We now calculate the rate of magnetic energy dissipation using the spectrum~(\ref{eq:spectrum}):
\be
-\frac{dE}{dt}=\eta \int\limits^{k_*}k_\perp^2 E(k_\perp)dk_\perp\propto S_L^{\frac{4}{7}\left(-\frac{3}{2}+\alpha\right)+\frac{6}{7}\left(3-\alpha\right)-1}. 
\label{eq:dissipation}
\ee
In a steady state, the rate of energy dissipation must be equal to the constant rate of energy cascade from the large-scale MHD turbulence independently of the Lundquist number. This defines the scaling of the energy spectrum uniquely: $\alpha=5/2$. 

%%%%
\paragraph{Discussion and Conclusion.}
The results derived above present a compelling case for revisiting the mechanism of energy dissipation envisioned in  existing Kolmogorov-like theoretical models of MHD turbulence.  Because of progressively increasing eddy anisotropy at small scales \cite{boldyrev_spectrum_2006,chandran_intermittency_2015,mallet_statistical_2016}, at a sufficiently small scale the reconnection time becomes comparable to the eddy turnover time. Reconnection, we have argued, disrupts the eddies at that scale, and implies that no such eddies can form at smaller scales. The energy cascade from the large scales, where it is injected, to the smallest scales, where it dissipates, must therefore proceed through a new sub-inertial stage where reconnection and the resulting structures --- plasmoids and associated flows --- are key players. 

As has been noted in the past, the presence of the large-scale magnetic field and the Alfv\'enic time scale implies that the Kolmogorov first self-similarity hypothesis may not hold for MHD turbulence \cite[e.g.,][]{perez_etal2014}. In particular, the spectrum of MHD turbulence may depend not only, or not at all, on the Kolmogorov-like dissipation scale. The presented analysis provides physical arguments for the existence  of alternative scales (\ref{eq:lambda_coppi}) and (\ref{eq:diss_scale}) that play a crucial role in MHD turbulence. The dissipation scale (\ref{eq:diss_scale}) decreases faster with the Lundquist number than the Kolmogorov scale (\ref{eq:lambda_dissipation}). This means, for example, that in order for the numerical simulations of MHD turbulence to be resolved, their discretization scale should decrease faster than the Kolmogorov scale (\ref{eq:lambda_dissipation}) as the Lundquist number increases. This property of MHD turbulence has also been discussed in~\cite[][]{perez_etal2014}.  

%-------------------------------------------------------------------------------

\paragraph{Acknowledgments.}
In the advanced stages of preparation of this manuscript, we became aware that a similar calculation (in particular, the derivation of Eq.~(\ref{eq:lambda_coppi})) has concurrently and independently been performed by A. Mallet, A. Schekochihin, and B. Chandran. We thank the authors for sharing their results with us.
NFL was supported by the NSF-DOE Partnership in Basic Plasma Science and Engineering, award no. DE-SC0016215. SB is partly supported by the National Science Foundation under the grant NSF AGS-1261659 and by the Vilas Associates Award from the University of Wisconsin - Madison. 

%-----------------------------------------------------------------------------------

%\bibliography{master}
%merlin.mbs apsrev4-1.bst 2010-07-25 4.21a (PWD, AO, DPC) hacked
%Control: key (0)
%Control: author (8) initials jnrlst
%Control: editor formatted (1) identically to author
%Control: production of article title (-1) disabled
%Control: page (0) single
%Control: year (1) truncated
%Control: production of eprint (0) enabled
%

\end{document}